\documentclass[12pt,preprint]{aastex}

\newcommand{\fuse} {{\it FUSE\/}}

\newcommand{\hst}  {{\it HST\/}}

\newcommand{\kms}  {\rm ~km\,s$^{-1}$}
\newcommand{\lya}  {Ly\,$\alpha$}

\newcommand{\ergcmsA} {\rm\,erg\,cm^{-2}\,s^{-1}\,\AA^{-1}}

\newcommand{\wl} {$\lambda$}
\newcommand{\wll} {$\lambda\lambda$}
\newcommand{\hh} {H$_2$}

\newcommand{\heii} {{\ion{He}{2}}}

\newcommand{\cii} {{\ion{C}{2}}}
\newcommand{\ciii} {{\ion{C}{3}}}

\newcommand{\Ni} {{\ion{N}{1}}}
\newcommand{\nii} {{\ion{N}{2}}}

\newcommand{\ovi} {{\ion{O}{6}}}

\newcommand{\SiIV} {{\ion{Si}{4}}}

\newcommand{\siv} {{\ion{S}{4}}}

\newcommand{\pv} {{\ion{P}{5}}}
\newcommand{\ari} {{\ion{Ar}{1}}}

\newcommand{\feii} {{\ion{Fe}{2}}}

\newcommand{\teff} {$T_{\rm eff}$}

\newcommand{\etacar} {$\eta$\,Car}
\newcommand{\etaA} {$\eta$\,Car\,A}
\newcommand{\etaB} {$\eta$\,Car\,B}

\shorttitle{Hot Companion of $\eta$ Carinae}
\shortauthors{Iping et al.}

\begin{document}

\title{Detection of a Hot Binary Companion of $\eta$ Carinae\footnote{Based on observations made with the NASA-CNES-CSA Far Ultraviolet Spectroscopic Explorer. \fuse\ is operated for NASA by the Johns Hopkins University under NASA contract NAS5-32985.}}

\author{Rosina C. Iping\altaffilmark{2,3},  George Sonneborn\altaffilmark{2}, Theodore R. Gull\altaffilmark{4}, Derck L. Massa\altaffilmark{2,5}, \& D. John Hillier\altaffilmark{6}}

\altaffiltext{2}{Laboratory for Observational Cosmology, Code 665, NASA's Goddard Space Flight Center, Greenbelt, MD 20771; rosina@taotaomona.gsfc.nasa.gov, george.sonneborn@nasa.gov,  massa@taotaomona.gsfc.nasa.gov}
\altaffiltext{3}{Dept. of Physics, Catholic University of America, Washington, DC  20064}
\altaffiltext{4}{Laboratory for Exoplanets and Stellar Astrophysics, Code 667, NASA's Goddard Space Flight Center, Greenbelt, MD 20771; theordore.r.gull@nasa.gov}
\altaffiltext{5}{SGT, Inc., 7701 Greenbelt Rd., Suite 400, Greenbelt, MD 20770}
\altaffiltext{6}{Department of Physics and Astronomy, University of Pittsburgh, Pittsburgh, PA 15260; djh@rosella.phyast.pitt.edu}

\begin{abstract}
We report the detection of a hot companion of $\eta$ Carinae using high resolution spectra (905 - 1180 \AA) obtained with the  Far Ultraviolet Spectroscopic Explorer (\fuse) satellite. 
Observations were obtained at two epochs of the 2024-day orbit: 2003 June during ingress to the 2003.5 X-ray eclipse and 2004 April  several months after egress. These data show that essentially all the far-UV flux from \etacar\ shortward of \lya\ disappeared at least two days before the start of the X-ray eclipse (2003 June 29), implying that the hot companion, \etaB, was also eclipsed by the dense wind or extended atmosphere of \etaA. Analysis of the far-UV spectrum shows that \etaB\ is a luminous hot star. The \nii\ \wll1084-1086 emission feature suggests that it may be nitrogen-rich.  The observed far-UV flux levels and spectral features, combined with the timing of their disappearance, is consistent with \etacar\ being a massive binary system.

\end{abstract}


\section{INTRODUCTION}

The star $\eta$ Carinae (HD 93308), a prominent member of the Trumpler 16 (Tr16) association, is undoubtedly the most well-known and well-studied 
luminous blue variable (LBV).  It is possibly the  most massive star in the Galaxy and its unstable nature has attracted many investigators (see Davidson \& Humphreys 1997 for a review).
There is now strong evidence that $\eta$ Carinae is a binary.
A period of $\sim$2023 days is inferred from periodic spectral and light changes in the visual (Damineli 1996, Damineli et al. 2000) and the near IR (Whitelock et al.  1994, 2004). 
The X-ray light curve varies strongly with a $2024\pm2$ day period (Ishibashi et al. 1999; Corcoran 2005). Two X-ray eclipses (1998.0 and 2003.5) have been observed in which the X-ray flux dropped  to less than $<$15\% of its peak for $\sim$3 months.  The hard X-ray spectrum suggests a colliding-wind binary (Ishibashi et al. 1999; Pittard  \& Corcoran 2002; Corcoran 2005), with the dense stellar wind from the massive LBV  primary (\etaA) colliding with the higher velocity, lower density wind of a hot O-type secondary (\etaB) in a highly eccentric 5.54 yr orbit. The presence of a hot secondary is also inferred from photoionization modeling of the variability of doubly-ionized lines from the Weigelt blobs (Verner et al.  2005). 

 Until now \etaB\ has evaded direct detection because \etaA\ dominates  the systemic light from the infrared  through the UV longward of \lya.  Furthermore, \etaA 's
 dense  wind and dusty ejecta extinguish much of its own UV and visible luminosity, which drop dramatically below 1250 \AA\ (Hillier et al. 2001; Gull et al. 2005a).

In this Letter we report the first spectroscopic evidence and direct detection of \etaB\ using Far Ultraviolet Spectroscopic Explorer (\fuse) spectra from two epochs.
One is a series of large aperture (LWRS) far-UV (FUV) spectra obtained in 2003 June as \etacar\ approached and entered the X-ray eclipse. The second is a set of LWRS and narrow slit (HIRS) spectra obtained in 2004, six months after the end of the X-ray eclipse.   The LWRS observations of
\etacar\ also include two 11th magnitude B stars that lie close to \etacar, which, as we shall see, account
for about half of the observed FUV flux in the LWRS observations.  Luckily, during some of the LiF1 channel HIRS observations of \etacar,
we serendipitously obtained spectra of the B stars through the HIRS slit of the SiC1 channel.  
Comparison of the SiC1 HIRS observation and the eclipse ingress observation
enables us to disentangle the LWRS spectra and reveal the spectrum of
\etaB.  However, because the LWRS spectra contain the B stars, it
is necessary to employ a three-step process to establish that the LiF1 HIRS
spectrum of \etacar\ is dominated by the flux of \etaB, and not \etaA.

First, we demonstrate that the LiF1 HIRS spectrum of \etacar\ plus the SiC1 HIRS spectrum of the nearby B stars account for all FUV flux from
the system obtained through the LWRS aperture at nearly the same epoch.
This implies that the LWRS spectra are dominated by \etacar\ and the nearby B stars and that there is very little contribution from other
sources, such as the surrounding nebulosity.
Second, we show that the LWRS spectrum of \etacar\ at the start of the X-ray minimum
(when the secondary is expected to be eclipsed) is nearly all due to the B
stars and that \etaA\ contributes very little to the FUV flux of the
system.
Third, we show that the FUV spectrum outside of eclipse (when \etaB\
is expected to contribute to the flux) is much stronger and consistent with a source of higher effective temperature than \etaA.

\section{OBSERVATIONS}
FUSE observed \etacar\ was observed by \fuse\ multiple times between 2000 and 2004, using the   LWRS aperture (30\arcsec$\times$30\arcsec). 
This Letter utilizes a subset of these data (see Table \ref{etaobs}). The observations were processed with the
\fuse\ data processing and calibration pipeline (CalFUSE 3.0).  Spectra from the individual exposures were coadded
with the IDL program {\tt fuse\_register}\footnotemark \footnotetext{See \fuse\ data analysis link  at
http://fuse.pha.jhu.edu.}.

The \fuse\ instrument (Moos et al. 2000; Sahnow et al. 2000)  obtains spectra in four independent optical
channels: LiF1 and LiF2 (995 -- 1185 \AA), and SiC1 and SiC2 (900 -- 1100 \AA).  Each channel has two segments (e.g. LiF1a, LiF1b) that each cover $\sim$100 \AA. The relative alignment of the four channels is influenced by the
thermal history of the satellite in the days preceding a given observation. 
Thermal flexure in the instrument  often results in alignment shifts of 10-20\arcsec\ in the other three channels relative to LiF1, which is held fixed by the offset guiding system.  For HIRS aperture (1\farcs25$\times$20\arcsec) observations, this often means the target is only in LiF1.

The SiC2 (920-1100 \AA) HIRS aperture  was on \etacar\ for 6.9 ks of the 17.1 ks April 11 observation.  After scaling the flux by a factor of 5, the SiC2b HIRS spectrum is essentially identical to that of LiF1a HIRS, indicating that the SiC2 aperture was not perfectly centered on target.  SiC2 adds the 1082 -- 1100 \AA\ region, containing the important \nii\ \wll1084-86 and \heii\ \wl1085 lines.

\begin{figure}
\plotone{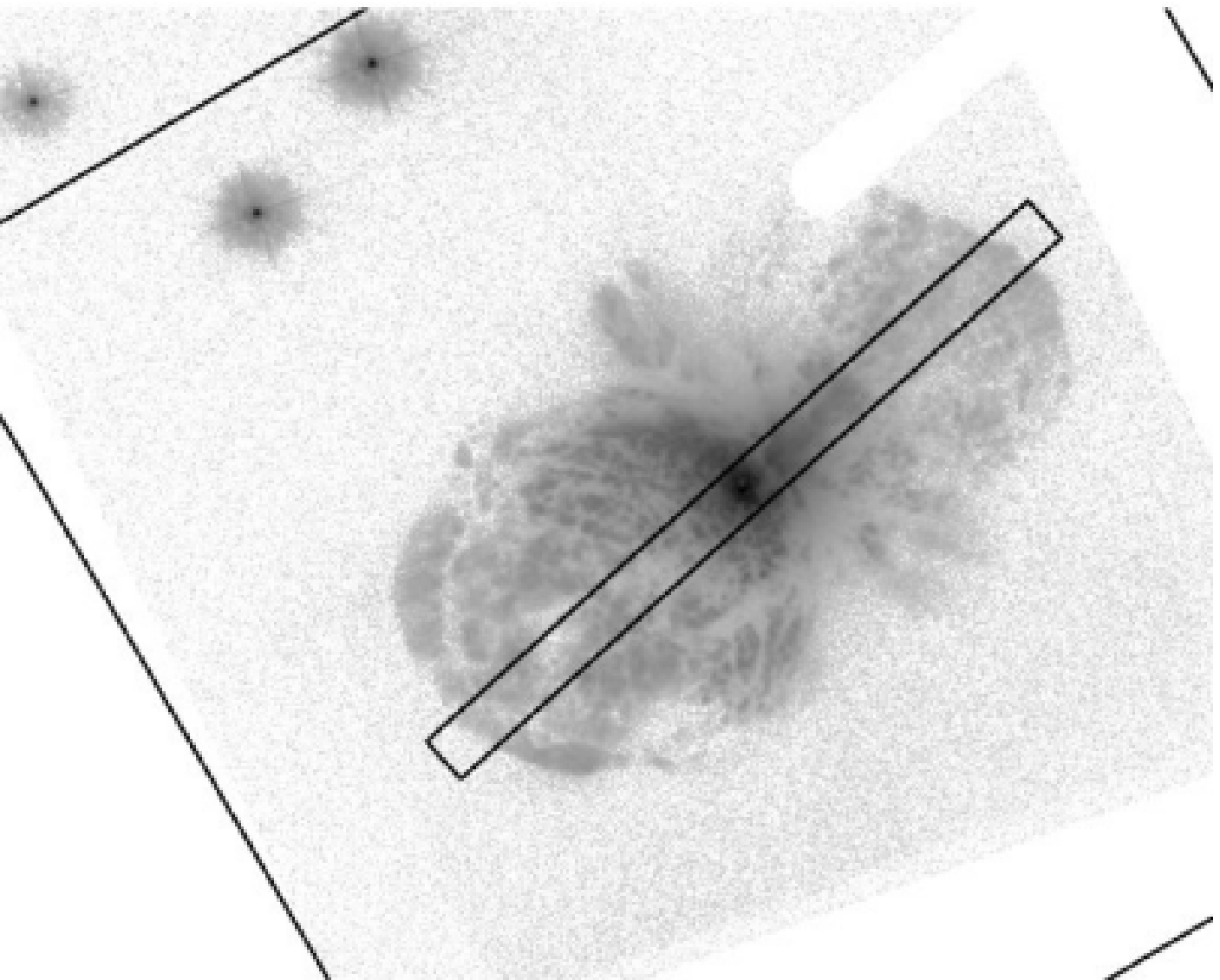}
\caption{\hst/ACS 2200 \AA\ (F220W) image of \etacar\ obtained on 2003 June 13.  The \fuse\ apertures are shown for 2004 March 30 (LWRS, 30\arcsec$\times$30\arcsec, p.a. = 120\arcdeg) and 2004 April 9-11 (LiF1a HIRS, 1\farcs 25$\times$20\arcsec, p.a. = 134\arcdeg). The stars Tr16-64, -65, and -66 are clearly visible. North is up, east is to the left.
\label{acsimage}}
\end{figure}

Given \etacar's spectroscopic variability and probable binary nature, we anticipated that its FUV flux would significantly decrease as it approached the X-ray eclipse.  
However, a surprising amount of FUV flux was present in all LWRS spectra of \etacar.   
 This resulted from spectral contamination of the LWRS observations by two 11th magnitude B stars, Tr16-64 and Tr16-65 (Feinstein et al. 1973), also members of Tr16, located just under 14\arcsec\ from \etacar\ (see Fig\,\ref{acsimage} and Table \ref{bstars}). They lie just inside the edge of the LWRS aperture at all position angles. A third B-type star, Tr16-66, is located 20\farcs 0 from \etacar\ and could, therefore, contribute to the LWRS  flux only in a narrow range of aperture position angles  (p.a. = 16\arcdeg$\pm$4\arcdeg, and increments of 90\arcdeg). Hubble Space Telescope( \hst) Advanced Camera for Surveys (ACS) near-UV images from 2003 June 13 provide astrometry and UV photometry for these stars. The 2200 \AA\ flux from Tr16-64 is $4.5\times 10^{-13} \ergcmsA$ and the Tr16-64:65:66 brightness ratio is 1.00:0.55:0.20 at 2200 \AA.  Tr16-64 has spectral type B1.5V (Levato \& Malaroda 1982). With Tr16 membership, similar $V$, visual and UV colors (implying no significant extinction differences), Tr16-65 must also be an early B star. Comparison of $V$ and UV brightnesses indicates that Tr16-66 would contribute less than 10\% of the combined FUV flux of Tr16-64 and Tr16-65 to LWRS observations.

Serendipitously, the aperture position angle and thermal alignment shifts during the LiF1 HIRS observations in 2004 April brought the SiC1 HIRS aperture to the location of Tr16-64 and TR16-65.  Examination of the time-tag data shows steady count rates in SiC1 HIRS for 48 out of 59 ks on April 9 and 10.  No signal from the B stars was present in the SiC1 HIRS channel on April 11. Although the exact SiC1, SiC2, and LiF2 aperture locations during the LiF1 HIRS observation are not known, there are no other stellar sources of FUV flux within the range of HIRS aperture motion.  The B-star spectrum recorded in SiC1 is a combination of Tr16-64 and Tr16-65.

\section{RESULTS}

The \fuse\ LiF1 HIRS spectrum gives the intrinsic FUV spectrum of \etacar, without contributions from Tr16-64 and Tr16-65.  Figure~\ref{lif1aplt}a shows the 1045\,--\,1090 \AA\ region of this observation and the SiC1a HIRS spectrum of Tr16-64 plus TR16-65 is shown in Figure~\ref{lif1aplt}b. Given the observed flux levels ($\sim2\times 10^{-12} \ergcmsA$), a significant fraction of the flux of these stars was in the SiC1 HIRS aperture. The B stars and \etacar\ share a rich interstellar spectrum that is characteristic of \fuse\ spectra of other Carina early-type stars (e.g., \hh, \cii, \feii, and \ari).  
Strong blue-shifted absorptions are also present in many transitions
in the FUV spectrum of eta Car. These include atomic lines also seen
in the interstellar medium (ISM) and Fe II absorption from levels
within 1000 cm$^{-1}$ of the ground state (Nielsen et al. 2005; Gull et
al. 2005a, 2005b).
These high-velocity features are seen in two principal groups ($-150$ \kms and $-450 \pm 80$ \kms). The stronger of these circumstellar features are marked in Fig.~\ref{lif1aplt}a as `CSM'. 
There is no evidence of high-velocity absorption in the \hh\ Lyman lines.

The SiC1 HIRS spectrum (Figure~\ref{lif1aplt}b) is characteristic of early main-sequence B stars (see Pellerin et al. 2002). The
\siv\ 1073 \AA line  shows only a photospheric line profile, that is no stellar wind feature in \siv\ 1062 or 1073, and indicates a low $v\sin i$ ($<$50 \kms).

Figure \ref{lif1aplt}c shows that the addition of the LiF1 HIRS and SiC1 HIRS spectra very accurately reproduces the LWRS spectrum of \etacar\ taken in 2004 March, indicating that the only significant sources of FUV flux in the LWRS aperture  are \etacar\ and the Tr16 B stars.  We now have the ability to extract the intrinsic FUV spectrum of \etacar\ from other LWRS observations by subtracting the SiC1 HIRS spectrum of Tr16-64 and Tr16-65.  

The point-source throughput of the HIRS aperture is $\sim 65 \pm 10$\% for targets centered in the aperture.  The flux for the HIRS spectra presented here have been corrected assuming a nominal 65\% throughput.  While this yields very satisfactory results, a uniform scaling of the HIRS fluxes may in fact not be a unique solution.

\begin{figure}
\plotone{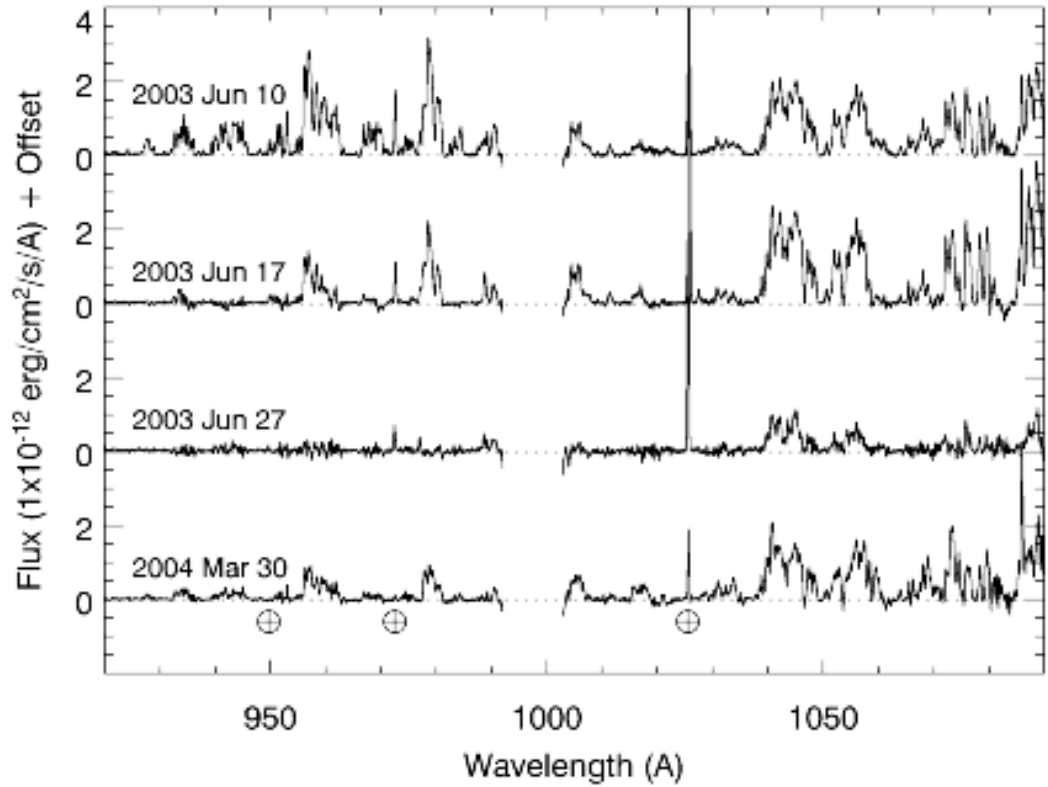}
\caption{\fuse\ LWRS aperture spectra (920--1090 \AA) of \etacar\ in 2003 and 2004 after subtracting the SiC1 HIRS spectrum of Tr16-64 and -65. Note that the net \etacar\ spectrum on 2003 June 27 is mostly consistent with zero flux, implying that the source of FUV radiation in \etacar\ has been eclipsed by \etaA.  The break in spectral coverage at 990--1002 \AA\  is the result of a gap between detector microchannel plates. \label{diffspec}}
\end{figure}

Several LWRS spectra of \etacar\ were obtained in 2003 June.  The X-ray flux was decreasing during this time frame as \etacar\ approached the X-ray eclipse.  The LWRS spectra on June 10 and 17 are nearly identical to spectra obtained in 2004 Mar.  The spectrum on 2003 June 27, however, is completely different, being nearly identical to the SiC1a HIRS spectra (Figure~\ref{lif1aplt}b).  The 2003 June and 2004 March LWRS spectra with the B stars subtracted are shown in Fig.~\ref{diffspec}.  There is almost a complete cancellation on June 27, with only a small amount of residual flux from \etacar\  present in limited wavelength regions (e.g. 1040--1046 \AA). 
In the 1100 - 1185\AA\ region, the June 27 spectrum has normal interstellar line profiles in many \feii\ lines, \Ni\ \wll1134, etc, (R. Iping et al.2005, in preparation). All the strong high velocity absorption present in every other LWRS observation of \etacar\ is not present, indicating that the primary source of FUV flux on that date is located outside of the \etacar\ Homunculus, that is the stars Tr16-64 and Tr16-65.

\section{DISCUSSION}
Our results show that essentially all the FUV flux from \etacar\ shortward of \lya\ disappeared at least two days before the start of the X-ray eclipse (2003 June 29, Corcoran 2005), implying that the source of the FUV flux was also eclipsed.  This conclusion is also supported by \hst Space Telecope Imaging Spectrograph (STIS) observations.
Examination of archival STIS FUV echelle spectra obtained with a 0\farcs 2$\times$0\farcs3 slit in 2003 show that above 1300 \AA\ there were increases in the strength of many absorption features on July 5 relative to earlier spectra.   From $\sim$1160 to 1200 \AA\ the spectrum was little changed on June 22 compared with earlier observations, but on July 5 this region had nearly zero flux.

The FUV flux shortward of 1000 \AA\ implies that \etaB\ has a higher \teff\ than \etaA\ (\teff=15,000K).  The Hillier et al. (2001, 2005) model of \etaA\ fades significantly below \lya\  and the flux from \etaB\ is expected to dominate \etaA\ only at these short wavelengths.  Verner et al. (2005) found that \etaB\  should have \teff $\sim$35,000 K in order to provide sufficient ionizing flux to excite Weigelt blob emission.  The \fuse\ spectrum is consistent with this result.

The \fuse\ wavelength region contains several features that should be key spectral diagnostics for the spectral classification and stellar wind properties of \etaB. These include \ovi\ \wll1032-38, \siv\ \wll1062-73, \ciii] \wl1176, \nii\ \wl1084-86, \SiIV\ \wll1122-28,  and \pv\ \wl1118-28. Few, if any, of these transitions are cleanly observed in the \fuse\ spectrum of \etacar, at least not in the form of profiles seen in other OB stars (see Pellerin et al. 2002 ; Willis et al. 2004). The strong, nearly saturated high-velocity \feii\ and \feii* absorption by Homunculus gas obliterates the \SiIV\ \wll1122-28 and \pv\ 1128 lines. The \ovi\ \wll1032-38 region does not have a clear P-Cygni line profile, but the general depression of the 1025-1038 \AA\ spectrum in \etacar\ relative to the B stars is characteristic of late O supergiants (e.g. HD 210809, O9 Iab, in Pellerin et al. 2002).

\begin{figure}
\plotone{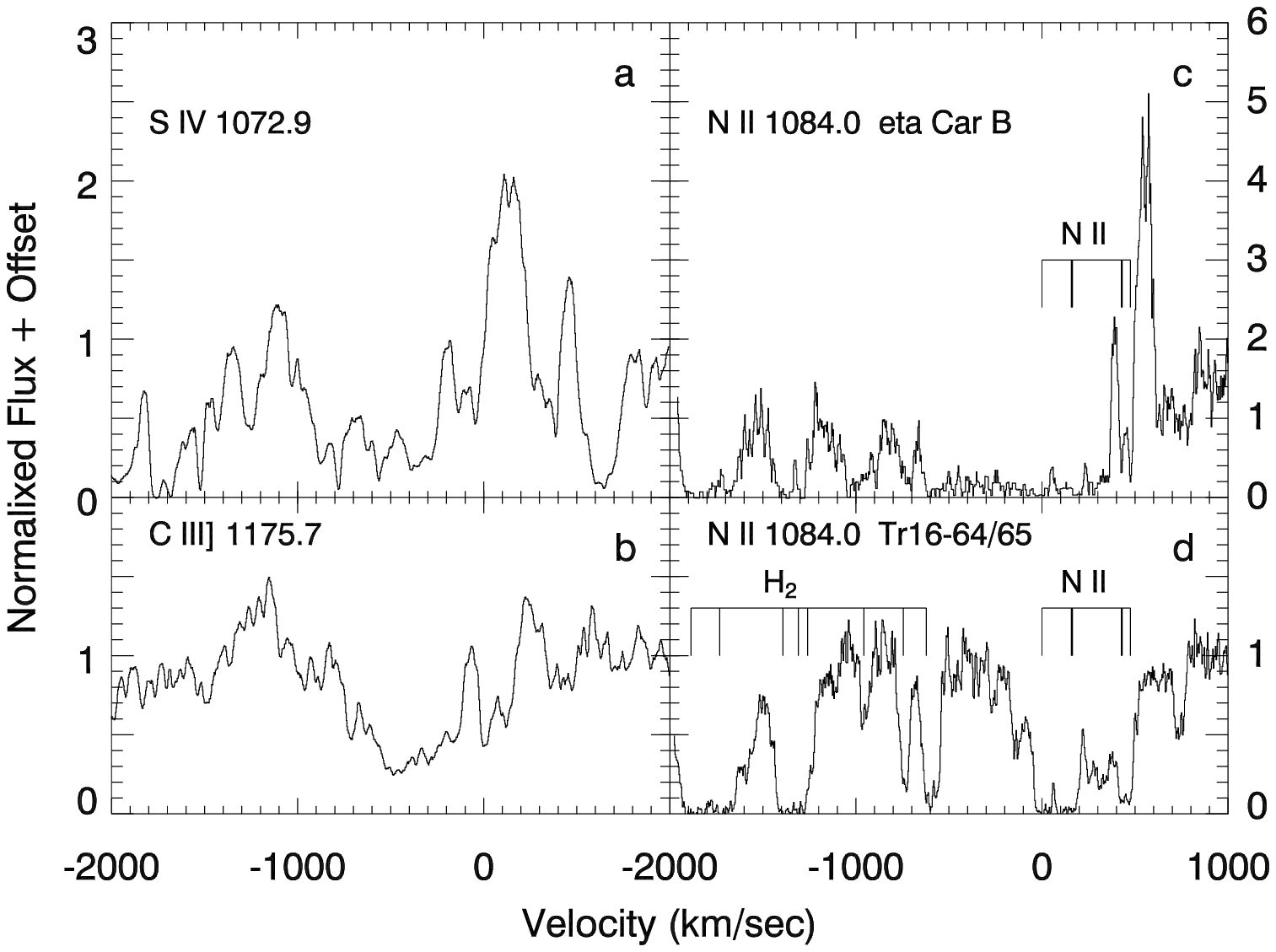}
\caption{Comparison of normalized line profiles of  a) \siv\ \wl1072.9, b) \ciii] \wll 1175, and c) \nii\ \wll1084-86 from the HIRS spectrum of \etaB.   Tr16-64/65 is shown in d) to illustrate the ISM features in the \nii\ region.   The spectra were normalized over a wide ($\sim10$ \AA) region around each feature.  The rest wavelengths for the ISM lines of the \nii\ multiplet and adjacent \hh\ are marked. The zero point of the velocity scale is the wavelength given in each panel. \label{windlines}}
\end{figure}

\siv, \ciii], and \nii\ line profiles (Fig.~\ref{windlines}) have broad absorption (600--1000 \kms) and red-shifted emission that is strong in \nii\ and weaker in \siv\  and \ciii].  \siv\ usually has a P-Cygni profile only in    O supergiants (Pellerin et al.  2002).  If formed in a wind of \etaB, the \siv\ and \ciii] features indicate a late-O spectral type.

\begin{figure}
\plotone{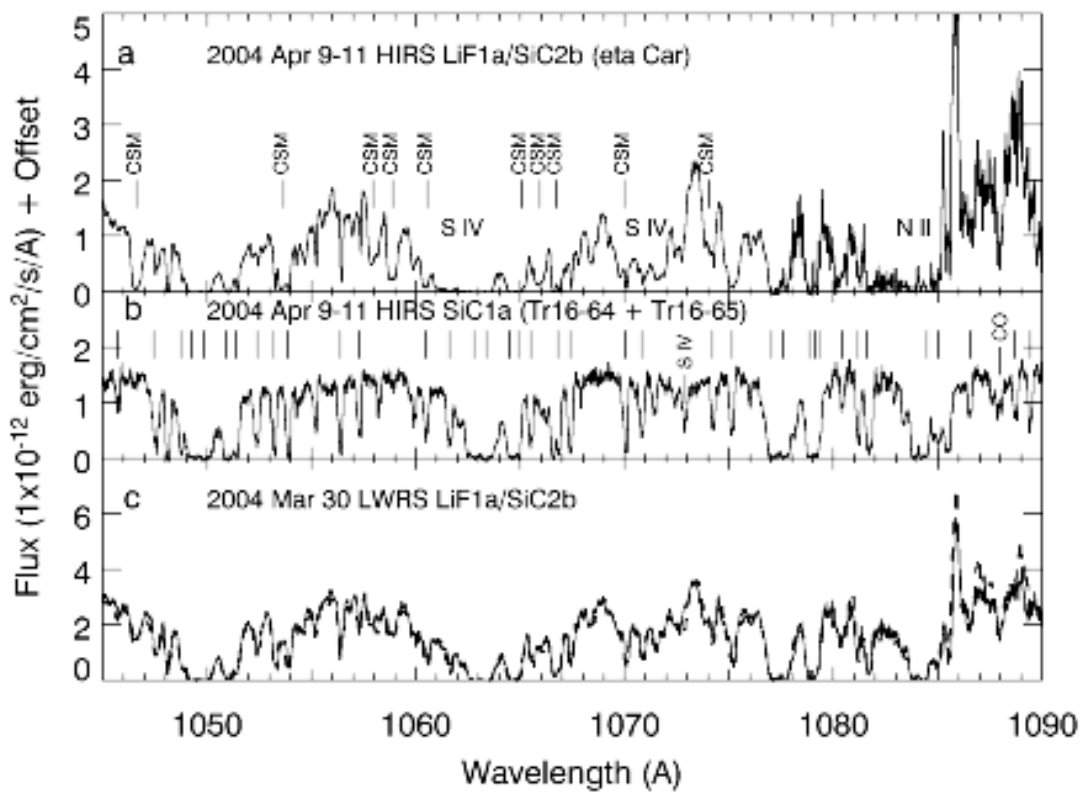}
\caption{a) \fuse\ LiF1a (1045--1077\AA) and SiC2b (1077--1090\AA) HIRS aperture spectra  of \etacar, corrected for HIRS point-source throughput of 65\%.  Locations of high-velocity circumstellar absorption are marked by `CSM'. The location of \siv\ 1062 -1073 are also indicated. b) SiC1a HIRS spectrum of Tr16-64 and Tr16-65. The vertical tick marks are the location of  H$_2$ Lyman lines with $J \leq 6$. c) \fuse\ LiF1a LWRS spectrum of \etacar\ (\etacar\ itself plus Tr16-64 and -65) from 2004 March 30. The dashed line is the sum of panels a and b, which almost exactly reproduces the LWRS spectrum. \label{lif1aplt}}
\end{figure}

The 1085 \AA\ region has a strong emission feature  at 1085.8 \AA\ (Fig.~\ref{lif1aplt}a, Fig.~\ref{windlines}).  This is the expected position for a P-Cygni feature from the \nii\ \wll1084-86 multiplet, and not emission from \heii\ \wl1084.9.  Significantly, this emission feature disappears in the June 27 observation, indicating that it is closely associated with \etaB, and not \etaA\ or the surrounding nebulosity.  The full width of the \nii\ emission is at least $\sim$ 130 \kms, although this is a lower limit because of strong interstellar absorption by \nii** \wl1085.55 and \wl1085.71 near their laboratory wavelengths (about -450 \kms\ in Fig.~\ref{windlines}c-d).  The saturated absorption is consistent with other transitions that have strong high-velocity features ($-100$ to $-600$ \kms) from the expanding ejecta of \etacar. 

A wind velocity of $\sim3000$ \kms\ for \etaB, as postulated in the Pittard \& Corcoran (2002) colliding winds model, is not evident in our data. The upper limit on the wind velocity for \etaB\ is $\sim$1100 \kms, based on the \siv\ \wll 1062 and 1073 and \ciii] \wll1175 lines. The profiles significantly overlap the CSM absorption velocities, making it more difficult to disentangle the two.  The higher wind velocity of the Pittard \& Corcoran model may not be observable if the wind of \etaB\ is highly distorted by the collision bowshock with the wind of \etaA, but this effect could be phase dependent in the eccentric orbit of the model.

\fuse\ spectra of \etacar\ and their timing relative to the X-ray light curve   demonstrate that there is a second, hotter star in the \etacar\ system.  The \fuse\ data do not conclusively identify the stellar type, but there are clues from several lines, including \nii\,\wll1084-86. This line is not observed as a P-Cygni profile in normal O stars (Pellerin et al. 2002, Walborn et al. 2002); it is \emph {only} seen in late-type nitrogen-rich O and Wolf-Rayet stars (see Walborn et al. 2002; Willis et al. 2004).  This raises the possibility that \etaB\ could be nitrogen rich, and possibly a Wolf-Rayet star.  Our upper limit  on the \etaB\ wind velocity is consistent with a late O/WR spectral type.  Alternatively, the \nii\ emission might arise from the colliding-wind shock interface and still be occulted during eclipse ingress.  
Better understanding of the \etacar\ system could come from integrated modeling of \etaB\ seen through the extended, expanding atmosphere and circumstellar material of \etaA.

\acknowledgements These observations were obtained for \fuse\ Cycle 4 Guest Investigator program D007. We thank the \fuse\ mission planning team for their excellent efforts to schedule these observations. This work was supported in part by NASA grant NAG5-12347 to Catholic University of America and contract S-67233-G to SGT, Inc. We thank Krister Nielsen for the spectroscopic analysis that helped identify the \feii\ high-velocity features and Nick Collins for providing the analysis of the ACS image.

\begin{deluxetable}{cllcrrr}
\tablewidth{0pt}
\tablecaption{\fuse\ Observation Parameters\label{etaobs}}
\tablehead{Obs. ID & Date & U.T. & $\phi$\tablenotemark{a} & $Expo (s)$ & Aper & p.a.}
\startdata
D007:01:02  & 2003 Jun 10 & 14:36 & 0.9888  & 15282  & LWRS   & 195\fdg 3 \\
D007:01:03  & 2003 Jun 17 & 03:33 & 0.9922  & 14311  & LWRS   & 201\fdg 1 \\
D007:01:04  & 2003 Jun 27 & 00:21 & 0.9972  &  4531  & LWRS   & 209\fdg 5 \\
D007:01:07  & 2004 Mar 29 & 14:36 & 0.1335  &  9202  & LWRS   & 119\fdg 4 \\
D007:01:08  & 2004 Mar 30 & 09:42 & 0.1340  & 34281  & LWRS   & 120\fdg 2 \\
D007:02:10  & 2004 Apr 9  & 03:23 & 0.1390  & 34476  & HIRS   & 131\fdg 3 \\
D007:03:11  & 2004 Apr 10 & 07:18 & 0.1395  & 24795  & HIRS   & 132\fdg 7 \\
D007:01:09  & 2004 Apr 11 & 21:16 & 0.1400  & 17118  & HIRS   & 134\fdg 5 \\
\enddata
\tablenotetext{a}{Phase defined by X-ray ephemeris 2450799.792 + 2024E (Corcoran 2005).}
\end{deluxetable}

\begin{deluxetable}{ccccccc}
\tablewidth{0pt}
\tablecaption{Stars Near \etacar \label{bstars}}
\tablehead{Name & $r$ (\arcsec) & p.a. & $V$\tablenotemark{a} 
& $(B-V)$\tablenotemark{a} & $(U-B)$\tablenotemark{a} & $F_{\lambda 2200}$\tablenotemark{b}}
\startdata
Tr16-64 & 13\farcs 97 & 41\fdg 0 & 10.72 & 0.10 & -0.74 & 4.5 \\
Tr16-65 & 13\farcs 84 & 60\fdg 3 & 11.09 & 0.14 & -0.65 & 2.5 \\
Tr16-66 & 20\farcs 00 & 61\fdg 0 & 11.98 & 0.16 & -0.57 & 0.9 \\
\enddata
\tablenotetext{a}{Data from Feinstein, Marraco \& Muzzio (1973)}
\tablenotetext{b}{UV flux from \hst/ACS F220W image; units are $10^{-13} \ergcmsA$}
\end{deluxetable}

\end{document}